\documentclass[a4paper,11pt]{article}

\usepackage[authoryear]{natbib}
\usepackage[utf8]{inputenc}
\usepackage[english]{babel}
\usepackage[T1]{fontenc}
\usepackage[a4paper,margin=1in]{geometry}
\usepackage{graphicx}
\usepackage{mathtools}
\usepackage{amsthm}
\usepackage{amssymb}
\usepackage{nicefrac}
\usepackage{booktabs,multirow}
\usepackage{array}
\usepackage{caption}
\usepackage{enumitem}
\usepackage{tikz}
\usepackage{xspace}
\usepackage[backref=page,hidelinks]{hyperref}

\usetikzlibrary{backgrounds}

\newcommand{\dist}{\mathsf{dist}}
\newcommand{\NP}{\text{NP}}
\newcommand{\R}{\mathbb{R}}
\newcommand{\calt}{\mathcal{T}\xspace}

\newtheorem{theorem}{Theorem}[section]
\newtheorem{lemma}[theorem]{Lemma}

\newtheorem{conjecture}[theorem]{Conjecture}
\theoremstyle{remark}
\newtheorem{remark}[theorem]{Remark}

\begin{document}

\title{Freeze-Tag is Strongly \NP-hard in 2D with \texorpdfstring{$L_p$}{Lp}
  Distances\thanks{Supported by São Paulo Research Foundation (FAPESP) grants \#2022/13435-4, \#2023/12529-8, and \mbox{\#2025/24919-0}, and by National Council for Scientific and Technological Development (CNPq) grant \#312271/2023-9.
  }}

\author{
  Lehilton Lelis Chaves Pedrosa%
  \thanks{\texttt{lehilton@ic.unicamp.br}; \url{https://www.ic.unicamp.br/~lehilton}.}
  \and
  Lucas de Oliveira Silva%
  \thanks{Corresponding author: \texttt{lucas.oliveira.silva@ic.unicamp.br}.}
}

\date{Institute of Computing, University of Campinas\\
  Campinas, São Paulo, Brazil}

\hypersetup{
  pdftitle={Freeze-Tag is Strongly NP-hard in 2D with Lp Distances},
  pdfauthor={Lehilton Lelis Chaves Pedrosa and Lucas de Oliveira Silva}
}

\maketitle

\begin{abstract}
  The Freeze-Tag Problem (FTP) asks for the minimum time needed to activate a swarm of robots, starting from a single active robot.
  When an active robot reaches a frozen robot, the latter becomes active; both robots may then move independently and activate further robots.
  We prove that FTP is strongly \NP-hard in the plane under every fixed rational \(L_p\) distance, \(1\le p<\infty\), and under \(L_\infty\).
  The geometric argument also applies to every fixed real \(p>1\) for which the metric admits an effective specification.
  For \(L_1\) and \(L_\infty\), the integer-coordinate decision problems are strongly \NP-complete.
  The reduction starts from Numerical 3-Dimensional Matching with distinct integers and also yields \NP-completeness for unweighted planar grid graphs.
\end{abstract}


\section{Introduction}

The Freeze-Tag Problem (FTP), introduced by~\cite{Arkin02}, models the automatic activation of a robot swarm from a single manually activated robot.
The problem gets its name from its similarity to the children's game also called ``freeze-tag''.
In the game, one player is designated as ``it'' and must tag other players to ``freeze'' them.
Meanwhile, unfrozen players work to ``unfreeze'' their frozen teammates by touching them.
FTP captures a simplified version of this game in which many players start frozen, and only one is unfrozen.
The objective is to minimize the time required to unfreeze every player.
No adversary can freeze players, so once a player is unfrozen, they never return to the frozen state.

FTP arises naturally in several applications.
Although initially formulated in the context of swarm robotics, it also models data transfer and network design problems~\citep{En19}.
Specifically, it is a variant of classical broadcast and peer-to-peer communication problems~\citep{Ra94}, with practical proximity constraints that may reflect technical or security requirements.
The problem is also closely related to multicast trees, communication structures commonly implemented in the application layer of computer networks~\citep{Pan98}.
Such trees are vital for geographically distributed applications that require efficient routing over the Internet.
An FTP solution is also a minimum-depth rooted binary tree spanning all robot positions in the input domain.
From this perspective, the problem can be seen as a variant of the Bounded-Degree Minimum Height Spanning Tree problem~\citep{Ju05}, also known as the Degree-Bounded Shortest Path Tree.
Similarly, it can be reduced to a version of the Bounded-Degree Minimum Diameter Spanning Tree problem~\citep{Kne05} with maximum degree three.

These connections highlight the mixed algorithmic nature of FTP.
As in minimum broadcast time, multicast, and gossiping problems, the goal is to disseminate information through a network~\citep{Ber73, He88, Ra94}.
Unlike these problems, FTP permits information transfer only through physical contact, so active agents must move before contacting others.
On the one hand, FTP involves routing: active robots must reach frozen robots before activating them or passing on information.
On the other hand, it can also be viewed as a cooperative variant of the Traveling Salesman problem, in which several agents become available over time and collectively visit all locations~\citep{Ar06}.
FTP is also connected to rendezvous search, in which mobile agents must meet to exchange information~\citep{Al03}.
From a network-design perspective, FTP is related to degree-bounded spanning trees~\citep{Kne05} and, more broadly, to tree spanners and metric tree embeddings~\citep{Cai95, Ba96, Fak04}.
Given its origins in swarm robotics, online variants are natural and have also been studied~\citep{Ha06}.

Despite these connections, FTP differs fundamentally from classical dissemination problems.
For example, although broadcast problems are often tractable on trees~\citep{Ra94}, FTP is already \NP-hard on trees~\citep{Arkin02}.
Depending on the underlying metric, FTP takes different forms.
For example, Arkin et al.~\citep{Arkin06} proved that FTP is \NP-hard on edge-weighted stars but polynomial-time solvable in the unweighted case.
Additionally, for general edge-weighted graphs, they showed that it is \NP-hard to obtain an approximation factor better than \(\nicefrac{5}{3}\), even for graphs of maximum degree four with a single robot per vertex.
On the algorithmic side, K{\"o}nemann et al.~\citep{Kne05} proposed an algorithm for degree-bounded minimum-diameter spanning trees with an approximation factor of \(O(\sqrt{\log n})\).
When the degree is limited to three, this algorithm applies to FTP and gives the best-known approximation guarantee for general edge-weighted graphs.

Beyond regular metric spaces, FTP has also been studied in an angular setting.
In the Angular Freeze-Tag Problem, introduced by~\cite{Fe18}, robots are fixed but equipped with rotating antennas, and they must point their antennas at one another to effectuate activations.
The turning angle is the cost.
This variant is \NP-hard to approximate within a factor better than \(\nicefrac{5}{3}\) already in the Euclidean plane, and only a \(9\)-approximation is known.

Recent work has also sought general upper bounds on the makespan of FTP instances in low-dimensional Euclidean spaces.
These works assume a normalized setting in which the source is at the origin, and all other robots lie within a unit ball centered at the origin.
It was shown by~\cite{BonichonDISC24} that there is a tight upper bound of \(5\) for the \(L_1\) distance in the Euclidean plane, which implies a bound of \(5\sqrt{2}\approx 7.07\) for the \(L_2\) distance.
Subsequently, Bonichon et al.~\citep{BonichonCCCG24} improved the bound for the \(L_2\) distance to \(4.63\).
Moving to three dimensions, Alipour et al.~\citep{AlipourAAMAS25} obtained upper bounds of \(13\) for \(L_1\) and \(22.52\) for \(L_2\).
Most recently, Alipour et al.~\citep{AlipourCCCG25} refined these results, obtaining bounds of \(4.31\) in the Euclidean plane under \(L_2\), and \(12\) and \(12.76\) in \(\R^3\) under the \(L_1\) and \(L_2\) distances, respectively.

Although~\cite{Arkin02} gave a PTAS for fixed-dimensional Euclidean spaces under every \(L_p\) distance, establishing hardness in low-dimensional geometric spaces has remained open.
Abel et al.~\citep{Yu17} proved that FTP is \NP-hard in the plane under the \(L_2\) distance, and Demaine and Rudoy~\citep{Erik17} subsequently established hardness in \(\R^3\) under every \(L_p\) distance with \(p>1\).
Pedrosa and de Oliveira Silva~\citep{Pedrosa23} closed the remaining three-dimensional case by proving strong \NP-hardness under \(L_1\), with rational coordinates polynomially bounded in the input size.
The complexity of planar \(L_p\) FTP for \(p\ne2\), including the Manhattan distance, nevertheless remained open.

The seminal FTP paper first raised the question of complexity in low-dimensional geometric spaces~\citep{Arkin02}.
They later conjectured that FTP is \NP-hard under both the $L_2$ and $L_1$ distances in 2D~\citep{Arkin06}.
The planar Manhattan case has been open for over two decades and is listed as Problem~35 of The Open Problems Project~\citep{TOPP}.
\begin{conjecture}[\cite{Arkin06}]
  FTP is \NP-hard in the Euclidean plane under both the \(L_1\) (Manhattan) and \(L_2\) (Euclidean) distances.
\end{conjecture}
Our result settles the remaining \(L_1\) case of this conjecture and, more generally, establishes hardness across the family of planar \(L_p\) metrics.

FTP has been studied more extensively through approximation algorithms in a variety of domains.
Table~\ref{state_of_the_art} summarizes the principal approximation and hardness results, including those proved here.
Further overviews are available in~\cite{Lu25, Sz03}.
\bigbreak

\renewcommand{\thempfootnote}{\arabic{mpfootnote}}
\noindent
\begin{minipage}{\textwidth}
  \centering
  \renewcommand{\arraystretch}{2}
  \newcolumntype{M}[1]{>{\centering\arraybackslash}m{#1}}
  \setlength{\tabcolsep}{2pt}
  \tiny

  \begin{tabular}
    {|M{20mm}|M{30mm}|M{32mm}|M{39mm}|M{30mm}|}
    \cline{1-5}
    \textbf{Version} & \textbf{Variant}                                                                                     & \textbf{Complexity} & \textbf{Approximation UB} & \textbf{Approximation LB} \\
    \cline{1-5}

    \multirow{3}{*}{\textbf{General graphs}}
                     & \multirow{2}{*}{weighted}
                     & \multirow{2}{*}{NPc~\citep{Arkin02}}
                     & \(O(\sqrt{\smash[b]{\log n}})\) and \(O\!\left(\frac{L}{D}\smash[b]{\log n}\right)\)\footnotemark[1]
                     & \multirow{2}{*}{\(\nicefrac{5}{3}\)~\citep{Arkin02}}                                                                                                                               \\
                     &                                                                                                      &                     & \citep{Kne05,Arkin03}     &                           \\
                     & unweighted with one robot per vertex
                     & NPc~\citep{Arkin03}
                     & \(22\)~\citep{Arkin03}
                     & \textbf{open}                                                                                                                                                                      \\
    \cline{1-5}

    \textbf{2D grid graphs}
                     & unweighted
                     & strongly NPc (Section~\ref{planar-lp})
                     & \(1+\varepsilon\)~\citep{Arkin02}
                     & not applicable                                                                                                                                                                     \\
    \cline{1-5}

    \multirow{2}{*}{\textbf{Trees}}
                     & weighted
                     & NPc~\citep{Arkin02}
                     & \(O(\sqrt{\smash[b]{\log n}})\)~\newline\citep{Kne05}
                     & \multirow{2}{*}{\textbf{open}}                                                                                                                                                     \\
                     & unweighted with one robot per vertex
                     & \textbf{open}
                     & \(22\)~\citep{Arkin03}
                     &                                                                                                                                                                                    \\
    \cline{1-5}

    \multirow{3}{*}{\textbf{Stars}}
                     & weighted with the same number of robots per leaf
                     & NPc~\citep{Arkin02}
                     & \(1+\varepsilon\)~\citep{Arkin02}
                     & not applicable                                                                                                                                                                     \\
                     & weighted
                     & NPc~\citep{Arkin02}
                     & \(14\)~\citep{Arkin02}
                     & \textbf{open}                                                                                                                                                                      \\
                     & unweighted
                     & P~\citep{Arkin02}
                     & not applicable
                     & not applicable                                                                                                                                                                     \\
    \cline{1-5}

    \textbf{Ultrametrics}
                     & weighted
                     & NPc~\citep{Arkin02}
                     & \(2^{O(\sqrt{\log\log n})}\)~\citep{Arkin02}
                     & \textbf{open}                                                                                                                                                                      \\
    \cline{1-5}

    \multirow{5}{*}{\textbf{Euclidean}}
                     & \(L_1\) or \(L_\infty\) distance in \(\R^2\)
                     & strongly NPc (Section~\ref{planar-lp})
                     & \multirow{5}{*}{\(1+\varepsilon\)~\citep{Arkin02}}
                     & \multirow{5}{*}{not applicable}                                                                                                                                                    \\
                     & fixed rational \(L_p\) distance in \(\R^2\), \(1<p<\infty\)
                     & strongly NPh (Section~\ref{planar-lp})
                     &                                                                                                      &                                                                             \\
                     & \(L_2\) distance in \(\R^2\)
                     & NPh~\citep{Yu17}
                     &                                                                                                      &                                                                             \\
                     & \(L_1\) distance in \(\R^3\)
                     & strongly NPc~\newline\citep{Pedrosa23}
                     &                                                                                                      &                                                                             \\
                     & \(L_p\) (\(p>1\)) distance in \(\R^3\)
                     & NPh~\newline\citep{Erik17}
                     &                                                                                                      &                                                                             \\
    \cline{1-5}

    \textbf{Angular}
                     & planar
                     & NPh~\newline\citep{Fe18}
                     & \(9\)~\newline\citep{Fe18}
                     & \(\nicefrac{5}{3}\)~\newline\citep{Fe18}                                                                                                                                           \\
    \cline{1-5}
  \end{tabular}

  \captionsetup{hypcap=false}
  \captionof{table}{Summary of approximation and hardness results for the Freeze-Tag Problem.}
  \label{state_of_the_art}
  \footnotetext[1]{\(L\) is the length of the longest graph edge, and \(D\) is its diameter.
    Weights are normalized so that the minimum weight is one.
  }
\end{minipage}

\paragraph{Our contribution}

We prove that FTP is strongly \NP-hard in \(\R^2\) under every fixed rational \(L_p\) distance for \(1\le p<\infty\), and under \(L_\infty\).
For \(p=1\) and \(p=\infty\), the corresponding integer-coordinate decision problems are strongly \NP-complete.
The same proof applies to every fixed real \(p>1\) for which exact distance comparisons can be effectively specified.

For every fixed \(p\), all locations have integer coordinates and polynomially bounded magnitude.
The \(L_1\) construction therefore also gives \NP-completeness on unweighted planar grid graphs.
Moreover, when schedules are represented rationally, planar \(L_1\) FTP is strongly ASP-hard; the same holds for the corresponding unweighted planar grid-graph instances.

\section{Preliminaries}%
\label{sec:preliminaries}

An FTP instance is a triple \((M,R,r_0)\), where \(M=(X,\dist)\) is a finite metric space, \(R\) is a multiset of points in \(X\), and \(r_0\in R\) is a distinguished occurrence.
The elements of \(R\) represent the robots' initial positions, and \(r_0\) is the unique robot that is active at time zero, called the \emph{source}.
All other robots are initially \emph{frozen}.
We treat the elements of \(R\) as labeled occurrences so that distinct robots may occupy the same point of \(X\).
For simplicity, an occurrence \(u\in R\) also denotes its position, and \(\dist(u,v)\) denotes the distance between the positions of robots \(u\) and \(v\).
Thus, distinct occurrences may have distance zero even though \(M\) itself is a metric space.

Once active, a robot moves at unit speed.
There are no collisions, congestion, or capacity constraints.
A frozen robot becomes active when an active robot reaches its position, and it may move immediately after activation.
All robots are otherwise identical and may coordinate their movements with complete knowledge of the input.
The objective is to activate every robot as early as possible.

For a geometric instance in dimension \(d\), \(X\) is a finite subset of \(\R^d\), and distances are induced by a fixed \(L_p\) norm.
For \(1\le p<\infty\), \(\|(x,y)\|_p=(|x|^p+|y|^p)^{\nicefrac{1}{p}}\), while \(\|(x,y)\|_\infty=\max\{|x|,|y|\}\); we write \(d_p(P,Q)=\|P-Q\|_p\).
We also use the standard graph-domain formulation for one consequence of the construction.
There, \(X\) is the vertex set of a connected edge-weighted graph, and \(\dist\) is its shortest-path metric.
An unweighted planar grid graph is a finite connected subgraph of the integer lattice in \(\R^2\), with edges between points at \(L_1\)-distance one.

A solution can be represented by a rooted tree \(\calt\), called a \emph{wake-up tree} or \emph{schedule}, with vertex set \(R\) and root \(r_0\).
The root has outdegree at most one, and every other vertex has outdegree at most two.
An edge \(uv\), directed away from the root, means that a robot available at \(u\) moves to \(v\) in time \(\dist(u,v)\).
The outdegree constraints reflect the number of available robots: before its first activation, the source supplies one moving robot; at a non-root activation, both the arriving and newly activated robots are available.

If several frozen robots occupy the same point, reaching that point activates them all simultaneously.
In a wake-up tree, we represent these simultaneous activations by an arbitrary binary tree of zero-weight edges between the corresponding occurrences.
All occurrences in this zero-weight tree have the same activation time.
The same convention applies to robots colocated with the source, which become active at time zero.

We use the physical interpretation of metric moves: if a shortest trajectory passes through a frozen occurrence, it activates that occurrence, which may then remain idle until needed.
This is equivalent to the wake-up-tree formulation.
A tree is realized by assigning the robots available at each activation to its child edges and following a shortest path.
Conversely, recording first activations, shortcutting traveled portions, and inserting zero-weight bookkeeping edges produces a tree with no larger makespan.

The \emph{activation time} of a robot \(v\) in \(\calt\) is the total edge weight on the unique path from \(r_0\) to \(v\).
The wake-up tree's \emph{makespan} is the maximum activation time, or equivalently the maximum weight of a root-to-leaf path in \(\calt\).
FTP asks for a wake-up tree of minimum makespan.
In the decision version, the input also contains a deadline \(L\), and the question is whether a wake-up tree of makespan at most \(L\) exists.

A solution is \emph{rational} if, at every time, each moving active robot follows a shortest path toward a distinct unclaimed frozen robot.
No two active robots claim the same frozen robot, and no superfluous movement is performed: once every frozen robot has been claimed, active robots without a target stop moving.
Every wake-up tree can be transformed into a rational solution without increasing its makespan~\citep{Arkin06}; hence, restricting attention to rational solutions is without loss of generality.

For \(1<p<\infty\), the exponent \(p\) is fixed in advance.
Our formal Turing-model theorem assumes a rational \(p\).
The same geometric proof applies to every fixed real \(p>1\) once an exact representation and comparison model for that metric is specified.

\section{Planar \texorpdfstring{$L_p$}{Lp} distances}%
\label{planar-lp}

For every fixed value of \(p\), the exponent in the coordinate bound below may depend on \(p\).

\begin{theorem}
  \label{thm:main}
  For every fixed rational $p\in[1,\infty)$, FTP is strongly \NP-hard in $(\R^2,d_p)$, even with integer robot locations and an integer deadline; the same holds for $p=\infty$.

  For $p\in\{1,\infty\}$, the integer-coordinate decision problem is strongly \NP-complete.
\end{theorem}

For $1<p<\infty$, we do not claim membership in \NP, thereby avoiding questions about exact comparisons of arbitrary sums of $L_p$ distances.

\subsection{Source problem: distinct-input N3DM}

We reduce from \emph{Numerical 3-Dimensional Matching with Distinct Integers} (N3DM).
An instance consists of three sets
\[
  U=\{u_1,\ldots,u_n\},\qquad V=\{v_1,\ldots,v_n\},\qquad W=\{w_1,\ldots,w_n\}
\]
of positive integers and a positive target $q \ge 3$, where all $3n$ integers in $U\cup V\cup W$ are unique.
The question is whether there exist permutations $\pi$ and $\sigma$ of $\{1,\ldots,n\}$ such that
\[
  u_i+v_{\pi(i)}+w_{\sigma(i)}=q \qquad\text{for every }i.
\]
This distinct-input restriction remains strongly \NP-complete~\citep{HWW08}.
It is also strongly ASP-complete~\citep{MITHardness24}.

Every yes-instance necessarily satisfies
\begin{equation}
  \label{eq:balance} \sum_{i=1}^n u_i+\sum_{j=1}^n v_j+\sum_{k=1}^n w_k=nq.
\end{equation}
It also satisfies
\[
  1\le u_i,v_j,w_k\le q-2,
\]
because two positive integers accompany every member of a target triple.
If either necessary condition fails, the reduction outputs any fixed no-instance of constant size.
We therefore assume \eqref{eq:balance} and these bounds throughout.

The distinctness assumption has two useful consequences for the reduction.
First, all constructed coordinates are distinct.
Second, no multiplicity bookkeeping is necessary.

\subsection{Wake-up trees and the forced backbone}

Along a root path, robot occurrences appear in a definite order.
The triangle inequality lower-bounds the path by the distances between any selected consecutive waypoints.
In particular, if a root path ending at $R$ contains occurrences $X$ and $Y$, then its length is at least
\begin{equation}
    \label{eq:rootpath} d(O,X)+d(X,Y)+d(Y,R) \quad\text{or}\quad d(O,Y)+d(Y,X)+d(X,R),
\end{equation}
according to which of $X$ and $Y$ occurs first.
We use this observation throughout.

Place the source at
\[
  O=(0,0)
\]
and a distinguished frozen terminal at
\[
  Z=(L,0),
\]
where $L$ is the deadline.
Place $n$ robots $a_i$ at distinct points
\[
  A_i=(\alpha_i,0),\qquad 0<\alpha_i<L.
\]
In our construction, the $A_i$'s are precisely the robot occurrences in the relative interior of the segment $OZ$, and no additional occurrence is colocated with $O$ or $Z$.

Call the unique root-to-$Z$ path in the wake-up tree the \emph{backbone}.
This path records the causal chain of activations that begins at the source and ends with the activation of $Z$.
Since the robots are identical, we may assume, without loss of generality, that the root robot performs every activation along the backbone.
Whenever a backbone occurrence is activated, the root robot continues toward the next backbone occurrence, while the newly activated robot is assigned to descendants outside the backbone.
Thus, the root robot traverses the entire backbone from $O$ to $Z$.
Each connected component obtained by deleting the backbone edges is therefore a rooted subtree attached to a backbone vertex; it records activations performed in parallel with the backbone traversal.

\begin{lemma}
  \label{lem:backbone}
  Assume that every shortest path from $O$ to $Z$ is the horizontal segment $OZ$.
  In every FTP wake-up tree of makespan at most $L$:
  \begin{enumerate} [label=(\roman*)]
    \item $Z$ is activated exactly at time $L$, and
    \item every occurrence at $A_i$ is activated at time $\alpha_i$.
  \end{enumerate}
\end{lemma}

\begin{proof}
  Since $d_p(O, Z)=L$, no wake-up tree can activate $Z$ before time $L$; feasibility therefore forces activation exactly at time $L$.
  Consider the backbone, traversed by the root robot under the convention above.
  Its total weight is at most $L$, while the triangle inequality gives the lower bound $d_p(O, Z)=L$.
  Thus, the backbone has weight exactly $L$ and realizes a shortest path from $O$ to $Z$.
  By hypothesis, it therefore follows the horizontal segment $OZ$.

  Consequently, the root robot reaches the $A$-points in increasing order, arriving at $A_i$ and activating it at time $\alpha_i$.
  No robot can reach $A_i$ earlier because $d_p(O,A_i)=\alpha_i$.
  Hence, every $A_i$ is activated exactly at time $\alpha_i$.
  Thus, the backbone contains every $A_i$, and every occurrence outside the backbone belongs to a rooted subtree attached to some backbone vertex, representing activations carried out in parallel after that vertex is reached.
\end{proof}

For $1<p<\infty$, the hypothesis of Lemma~\ref{lem:backbone} follows from strict convexity: equality in Minkowski's inequality forces all nonzero displacement vectors of a shortest polygonal path to be nonnegative scalar multiples of one another.
Hence, the shortest path from $(0,0)$ to $(L,0)$ is the horizontal segment.
For $p=1$, the length of a path equals the sum of the total variations of its coordinates.
A path of length $L$ therefore has a nondecreasing \(x\)-coordinate and a constant \(y\)-coordinate, so it is again the segment $OZ$.

\subsection{Exact construction for \texorpdfstring{$p=1$}{p=1}}

Set
\[
  a=2q,\qquad h_B=10q,\qquad b=(n+1)h_B+q,
\]
\[
  S=2a+2b+2q,\qquad h_C=10S,\qquad c=(2n+1)h_C+2q,
\]
and define
\begin{equation}
  \label{eq:L1deadline} L=c+S=2a+2b+c+2q.
\end{equation}
Place the source at $O=(0,0)$ and the backbone terminal at $Z=(L,0)$.
For every $u_i\in U$, create a robot at
\[
  A_i=(a+u_i,0),\qquad \alpha_i=a+u_i.
\]
The \(A_i\) are distinct because the \(u_i\) are distinct.

For every $v_j\in V$, create a robot at
\[
  B_j=\bigl(-jh_B,-(b+v_j-jh_B)\bigr),\qquad \beta_j=b+v_j.
\]
The quantity in the second coordinate is positive because
\[
  b+v_j-jh_B\ge h_B+q+1>0.
\]
Thus, every $B_j$ lies strictly in the southwest quadrant and $\|B_j\|_1=\beta_j$.

For every $w_k\in W$, create two labeled terminal robots $c_k^1,c_k^2$.
Let
\[
  \tau(k,s)=2(k-1)+s,\qquad s\in\{1,2\},
\]
and put
\[
  C_k^s=\bigl(\tau(k,s)h_C,c+2w_k-\tau(k,s)h_C\bigr), \qquad \gamma_k=c+2w_k.
\]
Both coordinates are positive, since \(\tau(k,s)\le2n\) and
\[
  c+2w_k-\tau(k,s)h_C\ge h_C+2q+2>0.
\]
Also $\|C_k^s\|_1=\gamma_k$.

All \(4n+2\) locations are distinct (see Figure \ref{fig:geometry} for an illustration).
The source, backbone points, \(B\)-points, and \(C\)-points are distinguished by their signs and by whether their second coordinate vanishes.
Within the three indexed families, distinctness follows respectively from the distinct $u_i$, the first coordinates $-jh_B$, and the distinct offsets $\tau(k,s)h_C$.
In particular, every wake-up-tree edge has positive length.

\begin{figure}[!h]
  \centering
  \begin{tikzpicture}[
      x=0.6cm,
      y=0.6cm,
      >=stealth,
      point/.style={
          circle,
          fill=black,
          inner sep=1.7pt
        },
      point label/.style={
          font=\small,
          inner sep=2pt
        },
      branch/.style={
          dashed,
          ->,
          thin
        }
    ]

    \coordinate (O) at (0,0);
    \coordinate (Z) at (12,0);

    \coordinate (A1) at (1.4,0);
    \coordinate (A2) at (3.1,0);
    \coordinate (A3) at (4.8,0);
    \coordinate (An) at (7.8,0);

    \coordinate (B1) at (-0.8,-2.5);
    \coordinate (B2) at (-1.7,-1.8);
    \coordinate (B3) at (-2.6,-1.1);
    \coordinate (Bn) at (-4.5,-0.5);

    \coordinate (C11) at (1.5,5.8);
    \coordinate (C12) at (2.6,4.8);
    \coordinate (C21) at (4.0,3.6);
    \coordinate (C22) at (5.1,2.6);
    \coordinate (Cn1) at (7.6,1.5);
    \coordinate (Cn2) at (8.7,0.7);

    \draw[->,thin]
    (-5.7,0) -- (12.9,0)
    node[right] {$x$};
    \draw[->,thin]
    (0,-4.1) -- (0,6.4)
    node[above] {$y$};

    \begin{scope}[on background layer]
      \draw[branch] (An) -- (B3);
      \draw[branch] (B3) -- (C11);
      \draw[branch] (B3) -- (C12);
    \end{scope}

    \node[point] at (O) {};
    \node[point label,above right] at (O) {$O$};

    \node[point] at (Z) {};
    \node[point label,above right] at (Z) {$Z$};

    \foreach \p/\lab in {
        A1/A_1,
        A2/A_2,
        A3/A_3,
        An/A_n
      }{
        \node[point] at (\p) {};
        \node[point label,above] at (\p) {$\lab$};
      }
    \node at (6.25,0.28) {$\cdots$};

    \foreach \p/\lab in {
        B1/B_1,
        B2/B_2,
        B3/B_3,
        Bn/B_n
      }{
        \node[point] at (\p) {};
        \node[point label,below left] at (\p) {$\lab$};
      }
    \node at (-3.55,-0.73) {$\cdots$};

    \foreach \p/\lab in {
        C11/C_1^1,
        C12/C_1^2,
        C21/C_2^1,
        C22/C_2^2,
        Cn1/C_n^1,
        Cn2/C_n^2
      }{
        \node[point] at (\p) {};
        \node[point label,above right] at (\p) {$\lab$};
      }
    \node at (6.35,1.95) {$\cdots$};

  \end{tikzpicture}
  \caption{Not-to-scale schematic of the construction, with the subtree corresponding to one triple shown by dashed arrows.}
  \label{fig:geometry}
\end{figure}

\subsubsection{Exact route lengths}
\label{sec:exactlen}

The quadrant choices imply
\begin{equation}
  \label{eq:L1AB} d_1(A_i,B_j)=\alpha_i+\beta_j
\end{equation}
and
\begin{equation}
  \label{eq:L1BC} d_1(B_j,C_k^s)=\beta_j+\gamma_k.
\end{equation}
Consequently, a route that starts at \(A_i\)'s activation time, visits $B_j$, and then reaches $C_k^s$ arrives at time
\begin{align}
  T(i,j,k,s) & =\alpha_i+d_1(A_i,B_j)+d_1(B_j,C_k^s)\notag \\  & =2a+2b+c+2(u_i+v_j+w_k)\notag               \\  & =L-2(q-u_i-v_j-w_k).
     \label{eq:L1route}
\end{align}
This route is feasible by time $L$ exactly when
\begin{equation}
  \label{eq:L1ineq} u_i+v_j+w_k\le q.
\end{equation}

\subsubsection{Terminal and separation properties}

\begin{lemma}
  \label{lem:L1terminal}
  Every $C$-occurrence is a leaf in every wake-up tree of makespan at most $L$.
\end{lemma}

\begin{proof}
  A terminal $C_k^s$ cannot be activated before time $\gamma_k=\|C_k^s\|_1$; after its activation, less than
  \[
    L-\gamma_k=S-2w_k<S
  \]
  time remains.

  Every other robot occurrence is at \(L_1\)-distance greater than \(S\) from $C_k^s$.
  If $C_k^s$ and $C_\ell^t$ are distinct and $d=\tau(k,s)-\tau(\ell,t)\ne0$, then
  \[
    d_1(C_k^s,C_\ell^t) =|d|h_C+|2(w_k-w_\ell)-dh_C| \ge2|d|h_C-2q \ge2h_C-2q>S.
  \]
  Moreover, $\tau(k,s)h_C>\alpha_i$ and $L>\tau(k,s)h_C$, so
  \[
    d_1(C_k^s,A_i)=\gamma_k-\alpha_i>S,
    \qquad
    d_1(C_k^s,B_j)=\gamma_k+\beta_j>S,
  \]
  and
  \[
    d_1(C_k^s,Z)
    =L+\gamma_k-2\tau(k,s)h_C
    \ge2h_C+S+4q+2>S.
  \]
  Thus, neither robot available at $C_k^s$ can reach another frozen occurrence before the deadline.
\end{proof}

\begin{lemma}
  \label{lem:L1backboneseparation}
  In every wake-up tree of makespan at most $L$, node $Z$ is a leaf, and no $B_j$ lies on the backbone.
\end{lemma}

\begin{proof}
  By Lemma~\ref{lem:backbone}, $Z$ is reached exactly at time $L$.
  Since all locations are distinct, every further activation would require positive time; hence $Z$ is a leaf.

  For every $j$ we have $d_1(O,B_j)=\beta_j$.
  Both coordinates of $B_j$ are negative, so
  \[
    d_1(B_j,Z)=L+\beta_j.
  \]
  Thus, if the backbone contains $B_j$, then it has length at least
  \[
    d_1(O,B_j)+d_1(B_j,Z)=L+2\beta_j>L,
  \]
  which is impossible.
\end{proof}

\begin{lemma}
  \label{lem:L1twoB}
  No root-to-$C$ path in the wake-up tree with makespan at most $L$ can contain two distinct $B$-occurrences.
\end{lemma}

\begin{proof}
  Suppose that the root-to-$C_k^s$ path contains two distinct occurrences, $B_j$ and $B_\ell$.
  Assume, without loss of generality, that $B_j$ precedes $B_\ell$ on this path.
  Since their first coordinates differ by at least $h_B$,
  \[
    d_1(B_j, B_\ell)\ge h_B.
  \]
  By applying the triangle inequality to the three corresponding
  sub-paths of the root-to-$C_k^s$ path, and using
  \eqref{eq:L1BC}, its total weight is at least
  \begin{align*}
    d_1(O,B_j)+d_1(B_j,B_\ell)+d_1(B_\ell,C_k^s)
     & \ge \beta_j+h_B+\beta_\ell+\gamma_k \\
     & \ge 2b+c+h_B+4                      \\
     & >2a+2b+c+2q=L.
  \end{align*}
  This contradicts the makespan bound.
\end{proof}

\subsubsection{Forced normal form}

\begin{lemma}
  \label{lem:normalform}
  Every wake-up tree of makespan at most \(L\) has the following normal form.
  There is a permutation $\pi$ of $\{1,\ldots,n\}$ such that, for every $i$, the child of $A_i$ outside the backbone is $B_{\pi(i)}$, and the two children of $B_{\pi(i)}$ are labeled $C$-occurrences.
  More precisely, each corresponding rooted subtree has the form
  \[
    A_i\longrightarrow B_{\pi(i)}
    \begin{cases}
      \longrightarrow C_{k_i}^{r_i}, \\[-2pt] \longrightarrow C_{\ell_i}^{s_i}.
    \end{cases}
  \]
  Moreover, every labeled $C$-occurrence appears as a child of exactly
  one $B$-vertex.
\end{lemma}

\begin{proof}
  The wake-up tree has $4n+2$ vertices and hence $4n+1$ edges, so its degree sum is $8n+2$.
  The source has degree $ 1$; the vertex $Z$ and the $2n$ occurrences $C_k^s$ are leaves and therefore have degree $1$; and every $A_i$ and every $B_j$ has degree at most $3$.
  Hence, the sum of these degree upper bounds is
  \[
    1+1+3n+3n+2n=8n+2.
  \]
  Since this equals the degree sum, every upper bound is attained.
  Consequently, in the wake-up tree, every $A_i$ and every $B_j$ has exactly two children, and the only leaves are $Z$ and the labeled $C$-occurrences.

  First, no edge joins two distinct \(B\)-vertices.
  Suppose, for a contradiction, that $B_j$ is the parent of $B_\ell$.
  Consider a leaf of the subtree rooted at $B_\ell$.
  If this leaf is $Z$, then $B_\ell$ lies on the backbone, contradicting Lemma~\ref{lem:L1backboneseparation}.
  If it is a $C$-occurrence, then the corresponding root-to-$C$ path contains both $B_j$ and $B_\ell$, contradicting Lemma~\ref{lem:L1twoB}.
  Hence, no two \(B\)-vertices are adjacent.

  It follows that every \(B_j\) has an \(A\)-vertex as parent.
  Indeed, its parent cannot be $Z$ or a $C$-occurrence because these vertices are leaves; it cannot be another $B$-vertex by the preceding paragraph; and it cannot be the source because the source has a unique child, namely the first $A$-occurrence on the backbone.

  Each $A_i$ has at most one $B$-child.
  Since $A_i$ lies on the backbone, one of its two children is its successor on the backbone, and Lemma~\ref{lem:L1backboneseparation} implies that this successor is not a $B$-vertex.
  Hence, the assignment sending each $B_j$ to its parent is a bijective map from the $B$-vertices to the $A$-vertices.
  Therefore, there is a permutation $\pi$ such that $B_{\pi(i)}$ is the unique \(B\)-child of $A_i$.

  Finally, both children of every $B_j$ are labeled $C$-occurrences.
  A child of $B_j$ cannot be the source, since the source is the root; it cannot be another $B$-vertex by the argument above, and it cannot be $Z$, because that would place $B_j$ on the backbone.
  It also cannot be an $A$-vertex, because every $A$-vertex lies on the backbone, which would again place $B_j$ there.
  Hence, both children of every $B_j$ are $C$-occurrences.

  The $n$ vertices $B_j$ have exactly two children each, giving exactly $2n$ distinct child vertices.
  Since there are exactly $2n$ labeled $C$-occurrences, every labeled $C$-occurrence is the child of exactly one $B$-vertex.
\end{proof}

\begin{lemma}
  \label{lem:balance-duplication}
  Let $\pi$ be a permutation of $\{1,\ldots,n\}$.
  For each $i$, suppose the two children of $B_{\pi(i)}$ are labeled $C$-occurrences whose associated numerical values are $x_i$ and $y_i$.
  Assume that, over all $i$, these $2n$ children comprise every labeled occurrence $C_k^1, C_k^2$ exactly once.
  If
  \[
    u_i+v_{\pi(i)}+x_i\le q, \qquad u_i+v_{\pi(i)}+y_i\le q
  \]
  for every $i$, then there exists a permutation $\sigma$ of $\{1,\ldots,n\}$ such that
  \[
    x_i=y_i=w_{\sigma(i)} \quad\text{and}\quad u_i+v_{\pi(i)}+w_{\sigma(i)}=q
  \]
  for every $i$.
\end{lemma}

\begin{proof}
  Summing the $2n$ inequalities gives
  \[
    2\sum_i u_i +2\sum_i v_{\pi(i)} +\sum_i(x_i+y_i) \le 2nq.
  \]
  Since $\pi$ is a permutation,
  \[
    \sum_i v_{\pi(i)}=\sum_j v_j.
  \]
  Moreover, because the children of the vertices $B_{\pi(i)}$ comprise
  every labeled occurrence $C_k^1,C_k^2$ exactly once, each value $w_k$
  appears exactly twice among
  \[
    x_1,y_1,\ldots,x_n,y_n.
  \]
  Hence,
  \[
    \sum_i(x_i+y_i)=2\sum_k w_k.
  \]
  Therefore, by~\eqref{eq:balance}, the left-hand side of the summed
  inequality is
  \[
    2\sum_i u_i+2\sum_j v_j+2\sum_k w_k=2nq.
  \]
  Thus, equality holds in the sum.
  Since each of the original inequalities has right-hand side $q$, each must be an equality.
  Consequently,
  \[
    u_i+v_{\pi(i)}+x_i=q \quad\text{and}\quad u_i+v_{\pi(i)}+y_i=q
  \]
  for every $i$, and therefore $x_i=y_i$.

  The elements of $W$ are pairwise distinct, so this common value determines a unique index $\sigma(i)$ satisfying
  \[
    x_i=y_i=w_{\sigma(i)}.
  \]
  Since each value $w_k$ occurs exactly twice among all the children of
  the $B$-vertices, and the two children of $B_{\pi(i)}$ already account
  for both occurrences of $w_{\sigma(i)}$, no distinct index $i'$ can
  satisfy $\sigma(i')=\sigma(i)$.
  Hence, $\sigma$ is injective.
  Because its domain and codomain both have cardinality $n$, it is a permutation.
  Finally, substituting $x_i=w_{\sigma(i)}$ into the corresponding equality gives, for every $i$,
  \[
    u_i+v_{\pi(i)}+w_{\sigma(i)}=q.
  \]
\end{proof}

\subsubsection{Correctness for \texorpdfstring{$p=1$}{p=1}}

Suppose first that the N3DM instance has permutations $\pi$ and $\sigma$ satisfying the target equalities.
Traverse the \(A\)-points in increasing distance from the origin and then continue to \(Z\).
When $A_i$ is activated, send its second robot to $B_{\pi(i)}$; the two robots then available at $B_{\pi(i)}$ travel to the two labeled copies of $C_{\sigma(i)}$.
Equation~\eqref{eq:L1route} places both arrivals exactly at time $L$.
Incidental earlier activations along these shortest trajectories do not affect feasibility, since the extra robots may remain idle until their prescribed use.

Conversely, consider a feasible wake-up tree and apply Lemma~\ref{lem:normalform}.
Let $x_i,y_i$ be the numerical $W$-values of the two terminal children of $B_{\pi(i)}$.
The root path to either child contains $A_i$ and $B_{\pi(i)}$, while Lemma~\ref{lem:backbone} gives the exact activation time $\alpha_i$ at $A_i$.
The root path lower bound \ref{eq:rootpath} and \eqref{eq:L1route} therefore imply
\[
  u_i+v_{\pi(i)}+x_i\le q, \qquad u_i+v_{\pi(i)}+y_i\le q.
\]
Lemma~\ref{lem:normalform} also ensures that these slots use every labeled terminal exactly once.
Lemma~\ref{lem:balance-duplication} now produces a permutation $\sigma$ satisfying
\[
  u_i+v_{\pi(i)}+w_{\sigma(i)}=q
\]
for all $i$.
This proves the equivalence of the N3DM and FTP instances.

The construction has $4n+2$ labeled robot occurrences.
Every coordinate has magnitude $O(n^2q)$, and all coordinates and the deadline are integers.
This proves strong \NP-hardness.
Since all \(L_1\) distances are integers, a wake-up tree is a polynomially verifiable certificate; hence the problem is strongly \NP-complete.

\subsection{The endpoint \texorpdfstring{$p=\infty$}{p=infinity}}

Define
\[
  \Phi(x,y)=(x+y,x-y).
\]
For every $(a,b)\in\R^2$,
\[
  \|\Phi(a,b)\|_\infty
  =\max\{|a+b|,|a-b|\}
  =|a|+|b|
  =\|(a,b)\|_1.
\]
The map $\Phi$ is a linear isometric bijection from the $L_1$ plane to the $L_\infty$ plane, with inverse
\[
  \Phi^{-1}(u,v)=\left(\frac{u+v}{2},\frac{u-v}{2}\right).
\]
Apply $\Phi$ to every point in the integer-coordinate \(L_1\) construction and retain the deadline.
Every physical trajectory and wake-up-tree edge preserve their lengths, so all activation times and makespans remain unchanged.
Integer coordinates remain integral, with magnitude increasing by at most a factor of two.
Strong \NP-completeness for $p=\infty$ follows.

\subsection{Approximate additivity for fixed \texorpdfstring{$1<p<\infty$}{1<p<infinity}}

Fix a rational constant $p\in(1,\infty)$.
The construction retains the horizontal backbone and the combinatorial mechanism forcing each \(A\)-vertex to activate one \(B\)-vertex, which in turn activates two \(C\)-occurrences.
The off-axis coordinates are redesigned so that the horizontal displacement along every intended path
\[
  A_i\longrightarrow B_j\longrightarrow C_k^s
\]
encodes the quantity $u_i+v_j+w_k$ exactly, up to a fixed scaling factor.
The transverse displacements only separate the constructed locations and contribute a nonnegative additive error.
Choosing the horizontal scales sufficiently large relative to these transverse displacements makes the total error on every intended path strictly less than one.
This preserves the integral gap between the cases $u_i+v_j+w_k\le q$ and $u_i+v_j+w_k>q$.

\subsubsection{A near-axis estimate}

\begin{lemma}
  \label{lem:nearaxis}
  For $x>0$ and $y\ge0$,
  \[
    0\le(x^p+y^p)^{ \nicefrac{1}{p}}-x\le\frac{y^p}{p x^{p-1}}.
  \]
\end{lemma}

\begin{proof}
  Put $t=(\nicefrac{y}{x})^p$.
  Then
  \[
    (x^p+y^p)^{ \nicefrac{1}{p}}=x(1+t)^{ \nicefrac{1}{p}}.
  \]
  The function $z\mapsto z^{
        \nicefrac{1}{p}}$ is concave on $(0,\infty)$, so its tangent at $1$ gives
  \[
    (1+t)^{
        \nicefrac{1}{p}}\le1+\frac{t}{p}.
  \]
  The lower bound is immediate.
\end{proof}

\subsubsection{Scales and locations}

Set $K=4$ and
\[
  r_p=\left\lceil\frac{p}{p-1}\right\rceil+1.
\]
Then
\begin{equation}
  \label{eq:rpexponent}
  r_p(p-1)>p.
\end{equation}
Define
\[
  a=2Kq,\qquad h_B=4Kq+4,
\]
\[
  b=(4nh_B)^{r_p}+a+Kq+1,
\]
\[
  S=2a+2b+2Kq+1,\qquad h_C=2S+1,
\]
\[
  c=(12nh_C)^{r_p}+2S+a+Kq+1,
\]
and set
\begin{equation}
  \label{eq:finitepdeadline} L=c+S=2a+2b+c+2Kq+1.
\end{equation}

Create the source at $O=(0,0)$ and the backbone terminal at $Z=(L,0)$.
For every $u_i\in U$, create a robot at
\[
  A_i=(a+Ku_i,0),\qquad \alpha_i=a+Ku_i.
\]
Because the $u_i$ are distinct, the $A_i$ are distinct.
For every $v_j\in V$, create a robot at
\[
  B_j=(-\beta_j,-jh_B),\qquad \beta_j=b+Kv_j.
\]
Thus, every $B_j$ lies in the southwest quadrant.
For every $w_k\in W$, create two labeled terminal robots $c_k^1,c_k^2$.
Let
\[
  \tau(k,s)=2(k-1)+s, \qquad s\in\{1, 2\},
\]
and put
\[
  C_k^s=(\gamma_k,\tau(k,s)h_C),\qquad \gamma_k=c+2Kw_k.
\]
Thus, every $C_k^s$ lies in the northeast quadrant.
All $4n+2$ locations are distinct.
The source, backbone points, $B$-points, and $C$-points are separated by their signs and by whether the second coordinate vanishes.
Within the three indexed families, distinctness follows respectively from the distinct $u_i$, the vertical coordinates $-jh_B$, and the distinct offsets $\tau(k,s)h_C$.
In particular, every edge of a wake-up tree has positive length, and all coordinates are integers.

\subsubsection{Uniform error bound}

Write
\[
  d_p(A_i,B_j)=\alpha_i+\beta_j+\varepsilon^{AB}_{ij}, \qquad d_p(B_j,C_k^s)=\beta_j+\gamma_k+\varepsilon^{BC}_{jks}.
\]
By Lemma~\ref{lem:nearaxis},
\begin{equation}
  \label{eq:ABerror} \varepsilon^{AB}_{ij} \le\frac{(nh_B)^p}{p(\alpha_i+\beta_j)^{p-1}} \le\frac{(nh_B)^p}{pb^{p-1}} <\frac{1}{p4^p}.
\end{equation}
Indeed, \(b\ge(4nh_B)^{r_p}\) and \eqref{eq:rpexponent} imply
\[
  b^{p-1}\ge(4nh_B)^{r_p(p-1)}>(4nh_B)^p.
\]
Since $h_C>h_B$ and $jh_B+\tau(k,s)h_C\le3nh_C$, the same calculation using $c\ge(12nh_C)^{r_p}$ gives
\begin{equation}
  \label{eq:BCerror}
  \varepsilon^{BC}_{jks}<\frac{1}{p4^p}.
\end{equation}
Consequently, for every \(A_i\), \(B_j\), and \(C_k^s\),
\begin{equation}
  \label{eq:totalerror} 0<\varepsilon^{AB}_{ij}+\varepsilon^{BC}_{jks} <\frac{2}{p4^p}< \frac{1}{2}.
\end{equation}
The final inequality holds for every $p>1$, and the middle expression tends to $\nicefrac{1}{2}$ as $p\downarrow1$.

\subsubsection{Approximate route lengths}

A route that starts at \(A_i\), visits \(B_j\), and then reaches \(C_k^s\) has arrival time
\begin{align}
  T(i,j,k,s) & =\alpha_i+d_p(A_i,B_j)+d_p(B_j,C_k^s)\notag \\  & =2a+2b+c+2K(u_i+v_j+w_k) +\varepsilon^{AB}_{ij}+\varepsilon^{BC}_{jks}\notag\\  & =L-1-2K(q-u_i-v_j-w_k) +\varepsilon^{AB}_{ij}+\varepsilon^{BC}_{jks}.
     \label{eq:finiteproute}
\end{align}
This route is feasible by time \(L\) exactly when
\begin{equation}
  \label{eq:finitepineq} u_i+v_j+w_k\le q.
\end{equation}
If \eqref{eq:finitepineq} holds, then \eqref{eq:totalerror} and \eqref{eq:finiteproute} give
\[
  T(i,j,k,s)<L-\nicefrac{1}{2}<L.
\]
Conversely, suppose that the route is feasible by time \(L\).
Combining \eqref{eq:finiteproute} with \eqref{eq:finitepdeadline} yields
\[
  2K\bigl(u_i+v_j+w_k-q\bigr) +\varepsilon^{AB}_{ij}+\varepsilon^{BC}_{jks}\le1.
\]
Since both error terms are nonnegative and their sum is less than one, we conclude that
\[
  2K\bigl(u_i+v_j+w_k-q\bigr)\le1.
\]
Now \(u_i+v_j+w_k-q\) is an integer.
If it were positive, it would be at least one, and the left-hand side would be at least $2K=8>1$, a contradiction.
Therefore, as in section \ref{sec:exactlen}, we obtain
\[
  u_i+v_j+w_k\le q.
\]

\subsubsection{Terminal and separation properties}

\begin{lemma}
  \label{lem:finitepterminal}
  Every $C$-occurrence is a leaf in every wake-up tree of makespan at most $L$.
\end{lemma}

\begin{proof}
  The separation argument is the same as in Lemma~\ref{lem:L1terminal}.
  Since $d_p(O,C_k^s)\ge\gamma_k$, less than
  \[
    L-\gamma_k=S-2Kw_k<S
  \]
  time remains after $C_k^s$ is activated.
  Every other occurrence is farther than \(S\), as follows from
  \begin{align*}
    d_p(C_k^s,C_\ell^t) & \ge h_C>S, & d_p(C_k^s,A_i) & \ge\gamma_k-\alpha_i>S, \\ d_p(C_k^s,B_j) & \ge\gamma_k+\beta_j>S, & d_p(C_k^s,Z) & \ge h_C>S,
  \end{align*}
  where the first bound concerns distinct $C$-occurrences.
  Hence, $C_k^s$ is a leaf.
\end{proof}

\begin{lemma}
  \label{lem:finitepbackboneseparation}
  In every wake-up tree of makespan at most $L$, node $Z$ is a leaf, and no $B_j$ lies on the backbone.
\end{lemma}

\begin{proof}
  As in Lemma~\ref{lem:L1backboneseparation}, Lemma~\ref{lem:backbone} and distinctness imply that \(Z\) is a leaf.
  Considering horizontal displacement alone gives
  \[
    d_p(O,B_j)+d_p(B_j,Z)\ge\beta_j+(L+\beta_j)>L,
  \]
  so no $B_j$ can lie on the backbone.
\end{proof}

\begin{lemma}
  \label{lem:finiteptwoB}
  No root-to-$C$ path in the wake-up tree with makespan at most $L$ can contain two distinct $B$-occurrences.
\end{lemma}

\begin{proof}
  We adapt the proof of Lemma~\ref{lem:L1twoB}.
  If $B_j$ precedes $B_\ell$ on a root-to-$C_k^s$ path, then the path leaves the backbone at some $A_i$ before reaching $B_j$.
  Considering only the horizontal displacement of the first and last edges and vertical separation between the two $B$-points, its length is at least
  \begin{align*}
    \alpha_i+d_p(A_i,B_j)+d_p(B_j,B_\ell)+d_p(B_\ell,C_k^s) & \ge2\alpha_i+\beta_j+\beta_\ell+\gamma_k+h_B \\  & \ge2a+2b+c+6K+h_B\\  & >2a+2b+c+2Kq+1=L,
  \end{align*}
  where the final inequality follows from $h_B=4Kq+4$.
\end{proof}

\subsubsection{Forced normal form}

\begin{lemma}
  \label{lem:finitepnormalform}
  Every wake-up tree of makespan at most $L$ determines a permutation $\pi$ of $\{1,\ldots,n\}$ such that, for every $i$, the child of $A_i$ that does not lie on the backbone is $B_{\pi(i)}$, and the two children of $B_{\pi(i)}$ are labeled $C$-occurrences.
  More precisely, the corresponding rooted subtree has the form
  \[
    A_i\longrightarrow B_{\pi(i)}
    \begin{cases}
      \longrightarrow C_{k_i}^{r_i}, \\[-2pt] \longrightarrow C_{\ell_i}^{s_i},
    \end{cases}
  \]
  Moreover, every labeled $C$-occurrence appears as a child of exactly one $B$-vertex.
\end{lemma}

\begin{proof}
  The proof of Lemma~\ref{lem:normalform} is combinatorial.
  It applies unchanged after replacing its three metric inputs by Lemmas~\ref{lem:finitepterminal}, \ref{lem:finitepbackboneseparation}, and \ref{lem:finiteptwoB}.
  Briefly, the degree sum
  \[
    2(4n+1)=8n+2=1+1+3n+3n+2n
  \]
  saturates all degree bounds; the two path-exclusion lemmas then rule out $B$-vertices on the backbone and $B$ to $B$ edges.
  Thus, every $B$ has a distinct $A$-parent and two $C$-children, yielding the claimed permutation and exhausting all labeled terminals.
\end{proof}

\subsubsection{Correctness for fixed \texorpdfstring{$1<p<\infty$}{1<p<infinity}}

For completeness, use the same wake-up tree as in the \(L_1\) construction: the backbone visits all \(A\)-points, each \(A_i\) sends its second robot to \(B_{\pi(i)}\), and the two robots there visit the two copies of \(C_{\sigma(i)}\).
For matching permutations, Equation~\eqref{eq:finiteproute} places every terminal arrival before \(L\), while the source, as it follows the backbone, reaches \(Z\) at time \(L\).

Conversely, Lemma~\ref{lem:finitepnormalform} supplies a permutation \(\pi\) and two numerical terminal values \(x_i,y_i\) for each \(B_{\pi(i)}\).
Applying Equation~\eqref{eq:finitepineq} to the two root paths gives
\[
  u_i+v_{\pi(i)}+x_i\le q, \qquad u_i+v_{\pi(i)}+y_i\le q.
\]
Since the slots exhaust all labeled terminals, Lemma~\ref{lem:balance-duplication} produces a permutation $\sigma$ satisfying
\[
  u_i+v_{\pi(i)}+w_{\sigma(i)}=q
\]
for all $i$.
This proves the equivalence of the N3DM and FTP instances.

For fixed $p$, the exponent $r_p$ is constant.
Since $h_B=O(q)$,
\[
  b=O((nq)^{r_p}).
\]
Also $h_C=O(b)$, and therefore
\[
  c=O\bigl(n^{r_p}(nq)^{r_p^2}\bigr).
\]
All integer powers have a constant exponent for fixed $p$, so the coordinates are computable by polynomial-time binary integer arithmetic.
On strongly bounded N3DM instances, every coordinate and the deadline have polynomial magnitude.
This proves strong \NP-hardness for every fixed rational \(1<p<\infty\).

\begin{remark}
  As $p\downarrow1$, the exponent $r_p$ grows on the order of $\frac{1}{p-1}$.
  The reduction is polynomial for every fixed $p$, but it is not a uniform polynomial-time reduction when $p$ is part of the input.
\end{remark}

\paragraph{Additional consequences.}

The \(L_1\) construction also yields strong \NP-hardness for FTP on unweighted planar grid graphs.
Indeed, all constructed locations have integer coordinates, and their pairwise $L_1$ distances are exactly the shortest-path distances in any sufficiently large rectangular unit grid containing them.
Thus, the same construction can be viewed as an FTP instance on this grid graph, with robots placed at the corresponding grid vertices.
Because the coordinate magnitudes are polynomially bounded on the strongly bounded N3DM instances used by the reduction, this grid graph has polynomial size.
Together with the integral-distance certificate for the \(L_1\) construction, this proves strong \NP-completeness.

Because distinct-input N3DM is strongly ASP-complete, as proved by \cite{MITHardness24}, this stronger hardness also carries over to FTP when solutions are rational.
By the normal-form lemmas, every feasible rational solution determines a unique permutation $\pi$ through the assignments $A_i\to B_{\pi(i)}$, and a unique permutation $\sigma$ through the two labeled $C_{\sigma(i)}$-occurrences activated from each $B_{\pi(i)}$.
Conversely, every N3DM solution \((\pi,\sigma)\) determines the corresponding rational solution.
Hence, the reduction preserves the solution set and transfers strong ASP-hardness to planar \(L_1\) FTP and to the corresponding grid-graph instances.
Through the \(L_1\)--\(L_\infty\) isometry, the same conclusion holds for planar \(L_\infty\) FTP.

\section{Conclusion}

We proved that the Freeze-Tag Problem is strongly \NP-hard in the plane under every fixed rational \(L_p\) distance for \(1\le p<\infty\), and under \(L_\infty\).
This settles the planar Manhattan case of the conjecture of~\cite{Arkin06}.
For \(L_1\) and \(L_\infty\), the integer-coordinate decision problems are strongly \NP-complete.

The exact \(L_1\) construction also gives \NP-completeness on unweighted planar grid graphs.
Because distinct-input Numerical 3-Dimensional Matching is strongly ASP-complete, the normal form also transfers strong ASP-hardness to planar \(L_1\) FTP and to the corresponding grid-graph instances.
Through the \(L_1\)--\(L_\infty\) isometry, the same conclusion holds for planar \(L_\infty\) FTP.
For fixed \(1<p<\infty\), the near-axis argument proves strong \NP-hardness, but we do not claim membership in \NP, since exact comparisons of arbitrary sums of \(L_p\) distances require further assumptions about the computational model.

Several algorithmic questions remain open.
Most notably, it is unknown whether FTP admits a constant-factor approximation for general edge-weighted graphs, or even for edge-weighted trees.
It would also be valuable to improve the best-known factor of \(22\) for unweighted general graphs with one robot per vertex.

\section*{Acknowledgments}

We thank Nicolas Bonichon, Nicolas Hanusse, and Ta{\"i}ssir Marc{\'e} for pointing out errors in an earlier version of our construction for $L_1$.

\clearpage
{
    \small
    \bibliographystyle{plainnat}
    \bibliography{bib}
}

\end{document}